\documentclass[prd, aps, superscriptaddress, twocolumn, preprintnumbers, floatfix, nofootinbib]{revtex4}

\usepackage{epsf}

\usepackage{amsmath}

\usepackage{amssymb}

\usepackage{graphicx}

\usepackage{dcolumn}

\usepackage{bm}

\def\be{\begin{equation}}
\def\ee{\end{equation}}
\def\bea{\begin{eqnarray}}
\def\eea{\end{eqnarray}}

\usepackage{color}

\begin{document}



\title{Cosmic Strings and the Origin of Globular Clusters}

\author{Alistair Barton, Robert H. Brandenberger and Ling Lin}

\email{alistair.barton@mail.mcgill.ca, rhb@physics.mcgill.ca, ling.lin2@mail.mcgill.ca}

\affiliation{Department of Physics, McGill University, Montr\'eal, QC, H3A 2T8, Canada}


\begin{abstract}

We hypothesize that cosmic string loops are the seeds about which globular
clusters accrete. Fixing the cosmic string tension by demanding that the
peak in the distribution of masses of objects accreting onto string loops
agrees with the peak in the observed mass distribution of globular clusters
in our Milky Way galaxy, we then compute the expected number density
and mass function of globular clusters, and compare with observations. Our
hypothesis naturally explains why globular clusters are the oldest
and most dense objects in a galaxy, and why they are found in the
halo of the galaxy.

\end{abstract}

\maketitle

\section{Introduction}

Globular clusters are spherical collections of stars which are found in the halos of galaxies
which orbit the galactic center. Globular clusters are older and more dense than other
star clusters found predominantly in the galactic disk. Our own Milky Way galaxy
contains about 150 known globular clusters. Their mass distribution (shown in Fig. 1)
has a peak at a mass of about $10^{5} M_{0}$, where $M_{0}$ is
the solar mass. The origin of globular clusters is still not well understood
(see e.g. \cite{GCreview} for a review).

In this Letter we propose that globular clusters are seeded by cosmic
string loops. Cosmic strings exist in many particle physics models
beyond the Standard Model (see \cite{CSrevs} for reviews on
cosmic strings and their role in cosmology). If Nature is described
by a theory which admits cosmic string solutions, then a network
of strings will inevitably form in the early universe and persist until
the present time \cite{Kibble}. The system of strings consists of
a network of infinite strings with a typical curvature radius $\xi(t)$ which 
scales with the Hubble radius $H^{-1}(t)$, where $H$ is the Hubble
expansion rate and $t$ is physical time, and a distribution of
string loops. At any given time, the distribution of string loops
is characterized by a {\it{critical radius}} $R_c(t)$ which dominates
the mass function of string loops.

The cosmological consequences of cosmic strings are characterized
by a single free parameters, namely the mass per unit length $\mu$
of the string (which is normally expressed in terms of the dimensionless
number $G \mu$, where $G$ is Newton's gravitational constant).
We will fix $G \mu$ by demanding that the mass which accretes
about a loop with radius $R_c$ agrees with the observed peak mass
in the globular cluster distribution. We then compute the predicted
number density of such loops, and find that it agrees with the
observed number density of globular clusters. We also compare
the mass function of globular clusters predicted by our model
with the data and obtain reasonable agreement. With one free
parameter we have therefore fixed the peak mass, the overall
number density and the mass function of globular clusters
\footnote{There is a slight caveat to this statement: the predicted
number density also depends on constants which are known
from analytical considerations to be of the order $1$ but whose
exact values have to be determined by numerical cosmic
string evolution simulations.}. 
Note that the accretion of matter
about string loops starts as soon as loops are created, i.e. much
earlier than the time of reionization. Hence, our mechanism offers
an explanation for why the globular clusters are the oldest components
of a galaxy. The string loops are initially distributed throughout
the region which eventually falls in to form the galaxy and will
hence end up in the galactic halo rather than the disk.

Note that in our current analysis, we are neglecting both the
initial translational velocities of the string loops as well as
the ``rocket effect'', an anisotropic gravitational radiation
reaction which can accelerate a string loop. 

In the following, we first review the distribution of cosmic string
loops resulting from any particle physics theory which gives
rise to strings. In Section 3 we present our computations. We fix
the one free parameter of the model, the string tension $G \mu$,
by demanding that the objects accreting around string loops
which dominate the loop mass function have the same mass
as those corresponding to the peak in the mass distribution of
globular clusters in the Milky Way. We then compute the predicted
overall number density and the mass function of objects  
accreting about such cosmic string loops and compare with
the observed mass distribution of globular clusters in our
Milky Way galaxy. Section 4 presents our conclusions.

\section{Cosmic String Loop Distribution}

Cosmic strings are one-dimensional topological defects which
arise in a large class of particle physics models beyond the
Standard Model. They are lines of trapped energy density,
and this energy gives rise to interesting signals in cosmology.
The important point \cite{Kibble} is that if Nature is
described by a particle physics model admitting cosmic
string solutions, then causality predicts that a network of
strings will inevitably form in the early universe and will persist to
the present time.

The energy scale $\eta$ of the new physics which yields
cosmic strings determines the string mass per unit length $\mu$.
In natural units the relation is $\mu = \eta^2$.
The amplitude of the gravitational effects of the strings
depends linearly on $\mu$. Strings come in two types:
a network of infinite strings with a correlation length
$\xi(t)$ which is proportional to the Hubble radius,
and a distribution of string loops. Long string
segments produce a conical discontinuity in space
which leads to characteristic signals in CMB temperature
maps \cite{KS}, polarization maps \cite{Holder1} and
21cm redshift maps \cite{Holder2} (see \cite{RHBCSreview}
for a recent overview of observational signals of long strings).

String loops can also play an important role in cosmology.
In fact, in early studies of the role of strings in
cosmology it was postulated \cite{CSearly} that strings
could account for all of the observed structure.
Specifically, it was postulated that galaxies in their
entirety were seeded by strings. The one string -
one galaxy hypothesis led to a value of the
string tension $G \mu \sim 10^{-6}$, which
corresponds to the energy scale $\eta$ of
particle physics ``Grand Unification''. However,
a cosmological model in which all fluctuations are
due to strings leads to incoherent perturbations and
predicts no acoustic oscillations in the CMB angular
power spectrum \cite{incoherent}. Thus, this
model is ruled out by the CMB data \cite{Boomerang}.
In fact, precision CMB observations such as ACT \cite{ACT},
SPT \cite{SPT}, WMAP \cite{WMAP} and Planck \cite{Planck}
lead to an upper bound on the string tension of  \cite{Dvorkin} 
\be \label{bound}
G \mu \, <  \, 2 \times 10^{-7}
\ee
(see \cite{older} for older results). Tighter bounds can possibly be
obtained by analyzing CMB maps in position space \cite{Danos}.
Constraints on the amplitude of the stochastic background
of gravitational waves produced by decaying cosmic string
loops lead \cite{Battye} to a stronger bound on $G \mu$ which, however,
depends sensitively on the ratio of loop production radius to
the Hubble radius (a number which the CMB bound is not
sensitive to).

As the discussion of the previous paragraph has shown,
cosmic strings cannot provide the dominant source of
fluctuations. The dominant source must be due to processes
in the early universe such as inflation \cite{Guth} or alternatives
(see \cite{ RHBrevAlt} for a review of alternatives such as
the ``Matter Bounce'' \cite{MB} or ``String Gas Cosmology'' \cite{SGC}).
Nevertheless, cosmic strings could have played an important
role in cosmology, as long as the tension obeys the bound (\ref{bound}).
For example, in many inflationary models cosmic strings will form
at the end of inflation \cite{Rachel}. Furthermore, any phase
transitions in matter which occur after the phase in the primordial
universe which produces the dominant source of fluctuations
(e.g. after the phase of inflation) will produce cosmic strings
as long as the particle physics model admits string solutions.
Hence, it is very interesting to search for the signals of strings
in cosmology. If the effects are not seen to the level of a particular
tension $\mu$, we will have ruled out all particle physics models
yielding strings 
with symmetry breaking scale $\eta$ larger than $\sqrt{\mu}$. If
strings are found, we can use them to explain cosmological
phenomena which up to now are not explained. It is a particular
application of the second possibility which we study in this Letter.
But before explaining our ideas we must return to the review of
the cosmic string scenario.

Cosmic string loops are typically generated by the intersection
of segments of the long strings, segments which move through
space at relativistic speeds. Causality tells us that the correlation
length of the string network (which gives the average separation
and average curvature radius of the long strings) is bounded from
above by the Hubble radius. Dynamical arguments show that
the correlation length at time $t$ cannot be parametrically smaller 
than $t$, otherwise string loops would be copiously produced
and $\xi(t) / t$ would increase. The network of long strings
thus takes on a ``scaling solution'' according to which the
statistical distribution of strings is independent of time if all
lengths are scaled to the Hubble radius.

The scaling distribution of strings implies that loops will be
formed at any time $t$ with a typical radius $R_f(t)$ given by
\be
R_f(t) \, = \, \frac{\alpha}{\beta} t \, ,
\ee
where $\alpha$ is a constant which must be determined from
numerical simulations. Recent simulations \cite{CSsimuls, Olum}
give $\alpha \sim 0.1$. The average length of a string loop
is $l = \beta R$, where $\beta$ is a constant which would be
$2 \pi$ if the strings were exactly circular. We shall use
the value $\beta = 10$.

Once formed, the number density of loops of radius $R$ redshifts
because of the expansion of space. Loops also slowly decay by
emitting gravitational radiation \cite{gravrad}. Loops with
radius smaller than
\be \label{critical}
R_c(t) \, = \gamma G \mu t
\ee
will decay in less than a Hubble time. Here, $\gamma$ is
another constant which is determined from numerical simulations.
Its value is of order $\gamma \sim 10^2$. Combining these
results, we find that the number density $n(R, t)$ of string loops of
radius $R$ is given at times $t > t_{eq}$ (where $t_{eq}$ is
the time of matter-radiation equality) by
\be \label{distrib1}
n(R, t) \, \sim \, t^{-2} R^{-2}
\ee
for loops which form after $t_{eq}$ and
\bea \label{distrib2}
n(R, t) \, &=& \,   N \alpha^{5/2} \beta^{-5/2} t_{eq}^{1/2} t^{-2} R^{-5/2}   
\,\,\, {\rm for} \, R > \gamma G \mu t \nonumber \\
n(R, t) \, &=& \, N \alpha^{5/2} \beta^{-5/2} t_{eq}^{1/2} \gamma^{-5/2} (G \mu)^{-5/2} t^{-9/2} \nonumber \\
&=& {\rm const} \,\,\, {\rm for} \, R < \gamma G \mu t \, , 
\eea
for loops which form before $t_{eq}$. In the above, the constant
$N$ depends on the average number $\tilde{N}$ of long string segments per Hubbe
volume. Since two long string segments are required to form a loop,
we could expect $N$ to depend on the square of $\tilde{N}$.
We will be interested in values of $G \mu$ for which the 
critical radius (\ref{critical}) at the present time $t_{0}$ is smaller 
than $\alpha \beta^{-1} t_{eq}$. Such loops form before the time of equal 
matter and radiation, and their distribution is given by (\ref{distrib2}).

The total number density of strings is obtained by integrating
(\ref{distrib2}) over the radius. The integral is dominated by $R \sim R_c(t)$.
For times between $t_{eq}$ and the present time $t_{0}$ we obtain
\be \label{density}
n_{total}(t) \, \simeq \, N \alpha^{5/2} \beta^{-5/2} \gamma^{-3/2} 
z_{eq}^{-3/4} (G \mu)^{-3/2} t^{-3} \, ,
\ee
where $z(t)$ is the cosmological redshift at time $t$, and 
where the constant $5/3$ resulting from integrating over the loop distribution
(\ref{distrib2}) has been absorbed in the factor $N$.
Note, in particular, that the number density is larger for smaller values of
$G \mu$. Hence, for such values of $G \mu$ there will be many
string loops embedded in the region which collapses to form a galaxy
(the galaxy seed being given not by cosmic strings, but by the dominant
source of fluctuations). In fact, the accretion of matter about the galactic
center sweeps up string loops from a larger region of space.
Working in the Zel'dovich approximation \cite{Zel}, the local number
density of string loops inside a galaxy will be enhanced by a factor of
$F$. Accretion and virialization in each direction leads to the estimate
$F = 64$. 

In the following we will explore the possibility that the string loops are
the seeds for the globular clusters embedded in the galactic halo.

\section{Globular Clusters from Cosmic String Loops}

According to linear cosmological perturbation theory, accretion of matter 
about a cosmic string loop starts (modulo logarithmic growth at
earlier times) at $t_{eq}$. At that time, the mass function of string loops
is dominated by loops with radius
\be
R \, = \, \gamma G \mu t_{eq} \, .
\ee
Whereas these loops will have decayed by the present time, the objects
they seed will keep growing. The local number density of such objects
inside a galaxy is hence given by
\bea \label{localdens}
n_{local} \, &=& \, F N  \alpha^{5/2} \beta^{-5/2} t_{eq}^{1/2} t_{0}^{-2} R_c(t_{eq})^{-3/2} \nonumber \\
&=& \, F N  \alpha^{5/2} \beta^{-5/2} \gamma^{-3/2} z_{eq}^{3/2} (G \mu)^{-3/2} t_{0}^{-3} \, .
\eea
The mass which has accreted about these seed loops at the present time
\footnote{We are here assuming that the accretion continues to the present
time.} is given by
\bea \label{mass}
M(\gamma G \mu t_{eq}, t_0) \, &=& \, \beta \gamma (G \mu)^2 z_{eq} \frac{t_{eq}}{G}
\nonumber \\
&=& \beta \gamma (G \mu)^2 z_{eq}^{-1/2} \frac{t_0}{G} \, .
\eea
Inserting the values of $t_{0}$ and $G$ we find
\be
\frac{t_{0}}{G} \, \sim \, 10^{23} M_{0} \, .
\ee
Making use of the values $\beta = 10$, $\gamma = 10^2$ and $z_{eq} = 10^4$ and setting
the mass (\ref{mass}) equal to the peak of the mass function of globular clusters in
our galaxy $M_{GC} \sim 10^5 M_{0}$ we find
\be \label{final}
G \mu \, \sim \, 10^{-9.5} \, .
\ee
Inserting this value into the local number density (\ref{localdens}), using (in addition
to the previously mentioned values for $\beta$ and $\gamma$) the values of the
constants (see \cite{Olum}) $\alpha = 0.3$, $N = 10^2$, and inserting the value of $t_{0}$
which yields $t_{0}^{-3} \sim 10^{-21} {\rm kpc}^{-3}$  we find
\be
n_{local} \, \sim \, 10^{-2} (kpc)^{-3} \, .
\ee
This is of the same order of magnitude as the observed number density of globular clusters
inside the Milky Way galaxy. 

The cosmic string loop seed hypothesis for global clusters predicts the primordial
mass function of globular clusters. The value of $M_c$, the peak in the mass function,
was used in our analysis to fix the string tension. However, once the tension is
fixed there is no more freedom in the mass distribution. For $M > M_c$ the
mass function scales as $M^{-5/2}$ (as follows directly from the string loop
distribution (\ref{distrib2})). For masses smaller than $M_c$ we predict a
linear decay. This follows from the fact that the loop radius distribution is
flat but loops with radius smaller than $\gamma G \mu t$ live only a fraction
of a Hubble time step which scales linearly with $R$. 

In Fig. 1 we compare
the predicted mass function (the solid line) with the observed distribution of
globular clusters in the Milky Way (histogram values). The values for
the blue histogram boxes come from the compilation in \cite{Gnedin}
of the properties of globular clusters in the Milky Way. The
theoretical curve is obtained by taking the comoving number density
$n(R, t_{eq}) z_{eq}^3$  of loops (where $n(R, t)$ is given in (\ref{distrib2})),
as a function of $R$, multiplying the result by the concentration 
factor $F$, allowing each loop to grow in mass by a factor of
$z_{eq}$ (independent spherical accretion), and converting the distribution 
into a mass distribution $n(M, t_0)$, taking into account
the Jacobian of the transformation from $R$ to $M$. The result
is then multiplied by the bin size $\delta M$ and by the volume of
the Milky Way galaxy.  Taking the bin size to be $\delta M = f M$,
where $f$ is a number, we obtain for the peak number density bin
\bea
\delta N \, &=& \, N F f \alpha^{5/2} \beta^{-5/2} \gamma^{-3/2} z_{eq}^{3/2} \nonumber \\
&& \,\times (G \mu)^{-3/2} V t_0^{-3} \,
\eea
where $V$ is the volume of the galaxy. The peak mass of the theoretical
curve is given by (\ref{mass}). In Fig. 1 the horizontal
axis is mass (logarithmic scale) while the vertical axis gives the number of
globular clusters (linear scale) inside of the Milky Way galaxy. The
exact value of $G \mu$ for the theoretical prediction has been tuned
to give the lowest value of $\chi^2$ between the theory curve and the
data. Its value is $G \mu = 5.43 \times 10^{-10}$.

In Fig. 2 we show how the peak position and amplitude of the mass
function shift as $G \mu$ varies. As $G \mu$ decreases, the peak mass
decreases but the predicted number of globular clusters increases. The
surprising result is that a good fit of both amplitude and peak number
density occur for the same value of $G \mu$ \footnote{To get the
fit of Figures 1 and 2 we have, however, used a value of $\alpha$ which
is slightly larger than that obtained from the simulations of \cite{Olum}.}.

Our analysis does not take into account interactions between globular clusters
and other processes which can change the mass of a globular cluster as a
function of time during the highly nonlinear process of galaxy formation.
Another effect which we have neglected in the current analysis is that
of initial translational motion of string loops.

\begin{widetext}

\begin{figure}
\includegraphics[width=15cm, height=12cm]{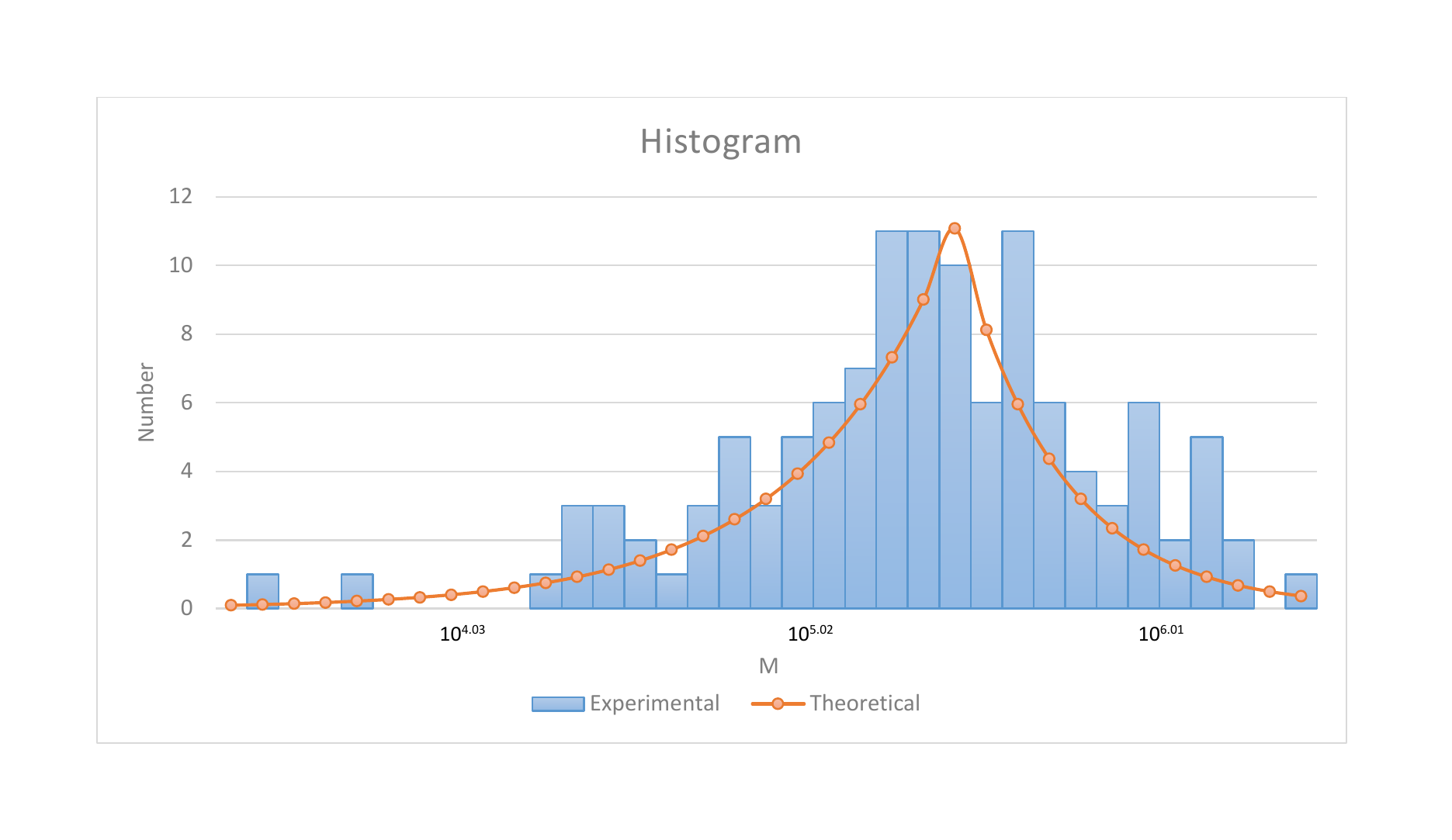}
\caption{Mass function of globular clusters in the Milky Way galaxy.  
The horizontal axis is mass on a logarithmic
scale, the vertical axis gives the number density on a linear scale.
The histogram show the data taken from \cite{Gnedin}. The solid curve is the
prediction of our model for the value of $G \mu = 5.43 \times 10^{-10}$ 
which minimizes $\chi^2$. The cosmic string parameters chosen are
described in the text. Note, in particular, that the overall amplitude 
of the theoretical curve is predicted once $G \mu$ has been
fixed to fit the peak location.} \label{globfig}
\end{figure}

\begin{figure}
\includegraphics[width=15cm, height=12cm]{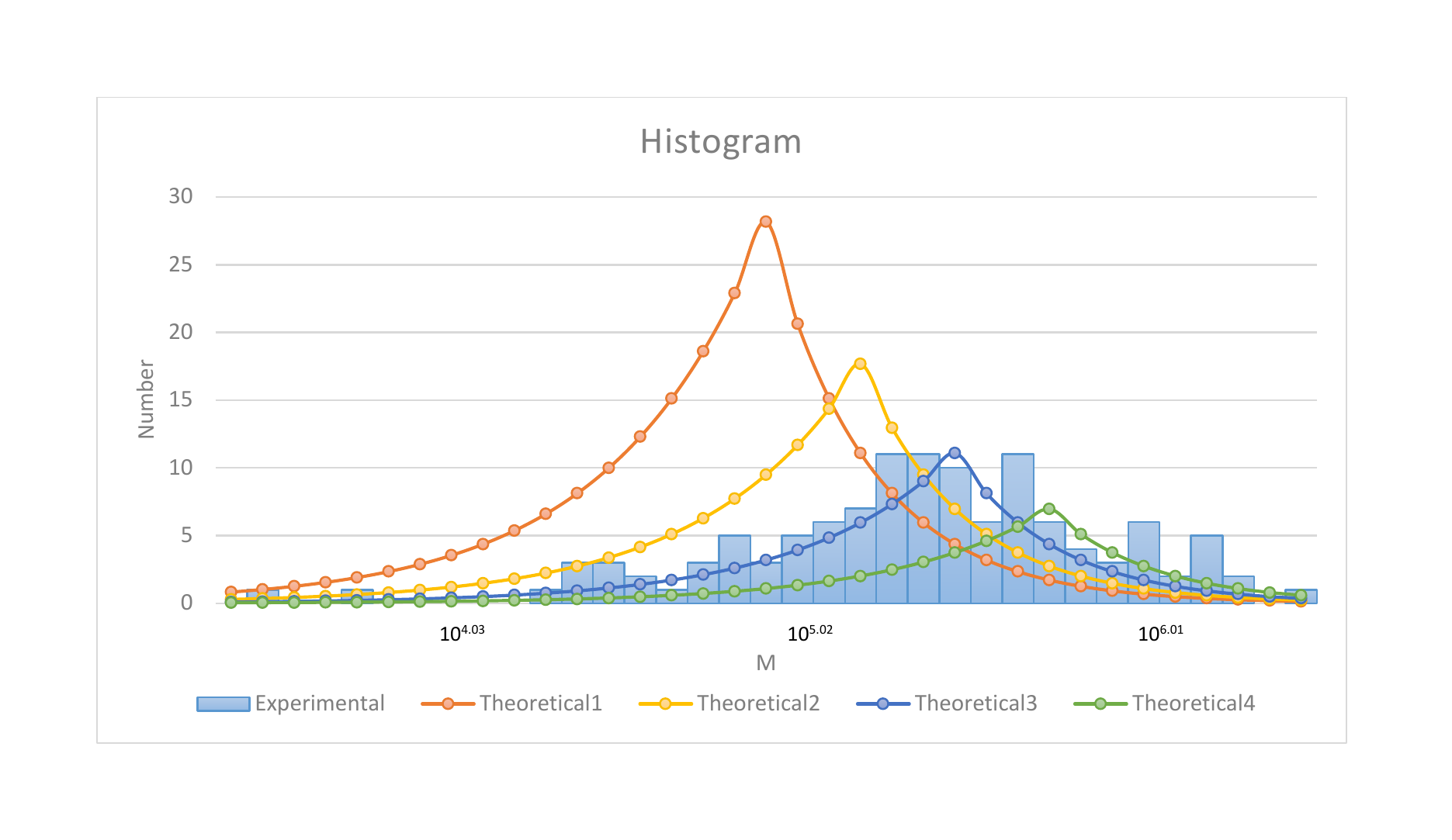}
\caption{Dependence of the mass function of our model on the
string tension $G \mu$. The curves shown are for $G \mu = 2.92 \times 10^{-10},
G \mu = 3.98 \times 10^{-10}, G \mu = 5.43 \times 10^{-10}$ and 
$G \mu = 7.41 \times 10^{-10}$ (in increasing
order of mass at the peak position). The axes and data
are the same as in the previous figure.}  
\end{figure}
\end{widetext}

\section{Conclusions and Discussion}

We have seen that both the number density and mass of string loops with a mass 
per unit length given by (\ref{final}) is correct to explain the origin of globular
clusters. We hypothesize that globular clusters are seeded by string loops which
dominate the loop mass distribution at the time $t_{eq}$ when the accretion of
cold matter onto the loops can start. We emphasize that fixing one free parameter
has allowed us to fit {\bf both} the mass and the number density of globular
clusters. In addition, the model predicts a mass function which is in reasonable
agreement with observations. Our model for the origin of globular clusters also 
naturally explains the fact that the globular clusters are the oldest star clusters 
in a galaxy, that they are the densest regions, and that they are distributed
in the galactic halo rather than confined to the galactic disk.

Our theory predicts the the number of globular clusters in a galaxy will scale
linearly with the mass of the galaxy. It also predicts that the mass function
of globular clusters should be universal across galaxies. Since the
energy distribution of dark matter falling onto a cosmic string is more peaked
than that falling into a Gaussian fluctuation \cite{CSHDM}, we predict falling 
velocity rotation curves about the center of a globular cluster.
Not all string loops are swept up into galaxies. Hence, our model also predicts
the presence of compact objects of globular cluster mass in the field between galaxies.
The corresponding number density does not include the factor $F \sim 10^2$.

In the current analysis we have neglected the initial translational motion of
the string loops. Since this peculiar velocity redshifts and since we are
considering string loops produced long before the time of equal matter and
radiation, the loop motion should be a small effect by the time that the loop
starts accreting matter at $t_{eq}$. We are currently in the process of working
out the details. An effect which might be more important (and which we are
neglecting here) is the rocket effect.

A question which we have not addressed here is the issue of star formation inside
the object which accretes onto the string loops (see e.g. \cite{Shlaer, Francis} for recent
studies of this question in the context of early structure formation from strings).
There may well be a significant difference in the star formation processes for
matter accreting onto a string loop inside a galaxy versus in the field.

\acknowledgements{We wish to thank G. Holder, Y. Omori, T. Webb 
and S. Yamanouchi for valuable
discussions. RB is supported by an NSERC Discovery Grant, and by funds 
from the Canada Research Chair program.}

\end{document}